\documentclass[prl,aps,twocolumn,showpacs,a4paper]{revtex4}
\usepackage{epsfig}
\date{\today}

\begin{document}

\title{Breakdown of integrability in a quasi-one-dimensional
ultracold bosonic gas}
\author{I.E. Mazets$^{1,2}$, T. Schumm$^1$ and
J. Schmiedmayer$^1$}
\affiliation{$^1$\, Atominstitut der \"Osterreichischen
Universit\"aten, TU Wien, A--1020 Vienna, Austria \\
$^2$\, A.F. Ioffe Physico-Technical Institute, 194021
St. Petersburg, Russia}

\begin{abstract}
We demonstrate that virtual excitations of higher radial modes in
an atomic Bose gas in a tightly confining waveguide result in
effective three-body collisions that violate integrability in this
quasi-one-dimensional quantum  system and give rise to
thermalization. The estimated thermalization rates are consistent
with recent experimental results in quasi-1D dynamics of ultracold
atoms.
\end{abstract}

\maketitle

Thermalization does not occur in integrable systems 
\cite{th1}, since the
number of their integrals of motion equals exactly  the number of
their degrees of freedom, thus such a system always ``remembers''
its initial state in the course of its dynamical evolution. In
an integrable system the finite spread of  initial energy
may lead only to  relaxation towards the generalized Gibbs (or
fully constrained thermodynamic) ensemble \cite{dn1}. Strictly
speaking, there is no thermalization in any {\em closed} system,
but for non-integrable systems the 
eigenstate thermalization hypothesis 
\cite{dn2}   holds, enabling  {\em dephasing} to
mimic the relaxation to the thermal equilibrium.

The Lieb-Liniger model \cite{LL} of spinless bosons with contact 
(point-like) interaction in 
one-dimension (1D) is a prime example of such an integrable system.

Ultracold atoms in strongly elongated  traps with $\omega _r \gg 
\omega_z$ ($\omega _r$, $\omega _z$ being the frequencies of the 
radial and longitudinal confinement, respectively) are an ideal 
system for studying 1D physics as long as both the temperature $T$ and
chemical potential $\mu$ are small compared to the energy scale
given by the transverse confinement:
\begin{equation}
\mu < \hbar \omega_r, \qquad k_BT<\hbar \omega_r . \label{eq:1}
\end{equation}

Strong inhibition of thermalization was observed in a beautiful
experiment with bosons deep in the 1D regime \cite{dw1}. However,
recent experimental results for a weakly interacting Bose gas
easily fulfilling the conditions of Eq. (\ref{eq:1})
\cite{js1,va,js2} are in a good agreement with the
thermal-equilibrium description of the 1D atomic ensembles.

In the present letter we investigate the breakdown of
integrability and thermalization in ultracold 1D bosons. The key
observation is that a radially confined atomic gas is never
perfectly 1D, and radial motion can be excited, either in reality
or virtually even if Eq. (\ref{eq:1}) holds. Therefore we call such
systems quasi-1D.

We start from identical bosons in a tight waveguide with radial
frequency $\omega _r$ ($\omega _z=0$), interacting via the
pseudopotential $4\pi \hbar ^2m^{-1}\alpha _s \delta ({\bf r}-{\bf
r}^\prime )$, where $m$ is the atomic mass, and $\alpha _s$ the
$s$-wave scattering length:
\begin{eqnarray}
    \hat{\cal H}_{3D}&=&\int d^3{\bf r}\,
    \left[ \hat{\psi }^\dag ({\bf r})
    \left( -\frac {\hbar ^2}{2m}\frac {\partial ^2}{\partial z^2}
    +\hat{H}^{(r)}  \right)
    \hat{\psi } ({\bf r})+ \right. \nonumber \\
    && \left.  \frac {2\pi \hbar ^2\alpha _s}m \hat{\psi }^\dag ({\bf r})
    \hat{\psi }^\dag ({\bf r}) \hat{\psi }({\bf r})\hat{\psi }({\bf r})
    \right] , \label{eq:2} \\
    \hat{H}^{(r)}&=&-\frac { \hbar ^2}{2m}\left(
    \frac {\partial ^2}{\partial x^2}+\frac {\partial ^2}{\partial y^2}
    \right) +\frac{m\omega _r^2}2 (x^2+y^2).   \label{eq:3}
\end{eqnarray}
We expand the atomic field operator as follows:
\begin{eqnarray}
    \hat{\psi }({\bf r})&=&\sum_{n,\ell ,k}\hat{a}_{ \{n,\ell \}\, k}
    \varphi _{n,\ell }(x,y)\frac {\exp (ikz)}{\sqrt{L}}.
    \label{eq:4}
\end{eqnarray}
Here $L$ is the quantization length and $\varphi _{n,\ell }(x,y)$
is the normalized eigenfunction of both the radial confinement
Hamiltonian, $\hat{H}^{(r)}\varphi _{n,\ell }(x,y)=(n+1)\hbar
\omega _r \varphi _{n,\ell }(x,y)$ 
and the $z$-projection of the
orbital momentum, $-i[x(\partial /\partial y)-y(\partial /\partial
x)] \varphi _{n,\ell }(x,y)=\ell  \varphi _{n,\ell }(x,y)$. The 
main quantum number $n=0,\, 1,\, 2,\, \dots \, $, and the 
orbital-momentum $z$-projection quantum number $\ell $ is restricted
by $|\ell |= \mathrm{mod}\, (n,2),\, \mathrm{mod}\, (n,2)+2,\,
\dots \, ,\, n-2,\, n$ and thus has the same parity as the main
quantum number. The atomic annihilation and creation operators
$\hat{a}_{ \{n,\ell \}\, k} $ and $\hat{a}_{ \{n,\ell \}\, k}
^\dag $ obey the standard bosonic commutation rules.

If two colliding atoms are initially in the transverse ground
state of the radial confinement (1D system), then their
orbital-momentum quantum numbers after collision are restricted to
$-\ell $ and $+\ell $.

For two-body collisions to contribute to thermalization, they have
to lead to transverse excitations.  This rate of population of the
radially excited modes by pairwise atomic collisions $\Gamma _{2b}$,
can be estimated for a non-degenerate Bose gas, using Fermi's
golden rule. For $k_BT<\hbar \omega _r$ this rate is
\begin{eqnarray}
    \Gamma _{2b} \approx \frac{2\sqrt{2} \hbar n_{1D}\alpha_s^2}{ml_r^3} \;
e^{-\frac {2\hbar \omega _r}{k_BT}} =
    2\sqrt{2} \omega _r \, \zeta \, e^{-\frac {2\hbar \omega _r}{k_BT}}, 
    \label{eq:20}
\end{eqnarray}
where $n_{1D}$ is the linear density and $l_r=\sqrt{\hbar /(m\omega
_r)}$ is the size of the transverse ground state. The
dimensionless quantity $\zeta =n_{1D}\alpha _s^2/l_r$ combines
two dimensionless parameters ($n_{1D}\alpha _s \propto
\frac{\mu}{\hbar \omega_r}$ and $\alpha _s/l_r$)
which can be seen as characterizing a 1D system \cite{foot2}.

Eq. (\ref{eq:20}) has a transparent physical interpretation:
$\Gamma _{2b}$ is related to the 3D atomic density, which is $\sim
n_{1D}/l_r^2$, times the $s$-wave scattering cross-section $\sim
\alpha _s^2$, times the exponential Boltzmann factor for the
fraction of atoms fast enough to scatter into higher radial modes,
times the corresponding velocity of the collision, $\sim \hbar
/(ml_r)$.

Calculating the numbers for the data in the $^{87}$Rb experiments
\cite{js2}, $\alpha _s = 5.3~\textrm{nm}$, $n_{1D}=50~ \mu $m$^{-1}$,
$\omega_r /(2 \pi )= 3$ kHz, $T=30$ nK ($\zeta \approx
0.007$) one obtains a collision rate of $\Gamma _{2b} \sim 0.02 \,
s^{-1}$, at least one order of magnitude too
small for the time scale of the experiment.

\begin{figure}[t]
 \epsfig{file=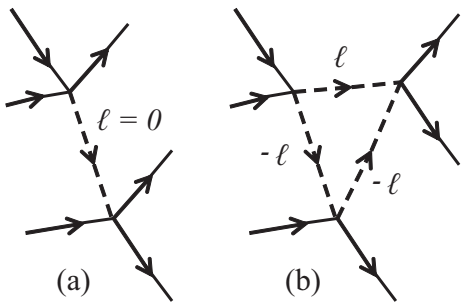,width=0.7\columnwidth}
 \caption {Feynman diagrams for the effective three-body processes
 in the second (a)
 and third (b) orders of perturbation theory. Solid and dashed
 lines correspond to atoms in the ground and excited states of the
 radial trapping Hamiltonian, respectively.}
 \label{fig:1}
\end{figure}

If the kinetic energy of the collision is less than $2\hbar \omega
_r$, then the radial modes can be excited only virtually and
contribute to the system dynamics only in the second and higher
orders of perturbation theory. If after the collision the radial
motion state is $| \{ n_1^\prime ,\ell _1^\prime \} , \, \{
n_2^\prime , \ell _2^\prime \} \rangle =| \{ 0,0 \} , \, \{ 2p ,0
\} \rangle $, then only one more collision is enough to quench the
virtual excitation and return the system on the energy shell
[Fig. \ref{fig:1}(a)].
Such a process yields an effective three-body
collision already in the {\em second} order of perturbation
theory.

In contrast processes involving a virtual excitation to $| \{
n_1^\prime ,-\ell \} , \, \{ n_2^\prime ,+\ell \} \rangle $, $\ell
\neq 0$, shown in Fig.~\ref{fig:1}(b), contribute only in the {\em
third} order, and thus will be neglected.

The small parameter in our perturbation calculation is
$n_{1D}\alpha _s$. To avoid complications related to the
confinement-induced resonance in 1D scattering \cite{ol1} we
assume $\alpha _s\ll l_r$ \cite{foot2}.

In a first step we evaluate the matrix element
\begin{eqnarray}
    \langle  \{ 0,0 \} , \, \{ 2p ,0 \} |\delta (x-x^\prime )
    \delta(y-y^\prime )| \{ 0,0 \} , \, \{ 0,0 \} \rangle = \nonumber \\
    \left( {2^{p+1}\pi l_r^2} \right) ^{-1}
    \label{eq:5}
\end{eqnarray}
that corresponds to two atoms in the ground state of the incoming
channel, one atom remaining in the same state, but the other one
being excited to a state with zero orbital-momentum quantum number
and even main quantum number $n=2p$, $p=0,\, 1,\, 2,\, ... $ ($n$
and $\ell $ are required to have the same parity).

The result of Eq. (\ref{eq:5}) should not be confused with the
matrix element where the outgoing channel is characterized by
excitation of a higher radial mode of the {\em relative} motion of
two atoms as discussed in \cite{yur1o}, which equals to $(2\pi
l_r^2)^{-1}$.  The latter is a linear combination of the matrix
elements corresponding to vertices on {\em both} Fig.~\ref{fig:1}(a)
and Fig.~\ref{fig:1}(b), and thus can not be applied to the
calculation of the second-order process.

Using the matrix element Eq. (\ref{eq:5}) we can rewrite Eq.
(\ref{eq:2}) and by adiabatically eliminating the radially excited
mode operators obtain the effective 1D Hamiltonian:
\begin{eqnarray}
    \hat{\cal H}_{1D}&=&\hat{\cal H}_0+\hat{\cal H}_{1D}^{(3b)} , \label{eq:10} \\
    \hat{\cal H}_{1D}^{(3b)}&=&-\frac {2\xi \hbar \omega _r\alpha _s^2}L
    \sum \hat{a}_{k_1^\prime }^\dag \hat{a}_{k_2^\prime }^\dag
    \hat{a}_{k_3^\prime }^\dag \hat{a}_{k_1}\hat{a}_{k_2}\hat{a}_{k_3}
\label{eq:11}
\end{eqnarray}
where the summation in Eq. (\ref{eq:11}) is taken over all the
kinetic momenta obeying the conservation law $k_1 ^\prime + k_2
^\prime + k_3 ^\prime =k_1 + k_2 +k_3 $, $\hat{a}_k \equiv
\hat{a}_{ \{0,0\} \, k}$ and $\xi =4 \,
\mathrm{ln}\, (4/3) \approx 1.15 $. The relative contribution 
$(\xi -1)/ \xi $ of the virtual states with the
excitation energy higher than $2\hbar \omega _r$ is remarkably small.

\begin{figure}
  \epsfig{file=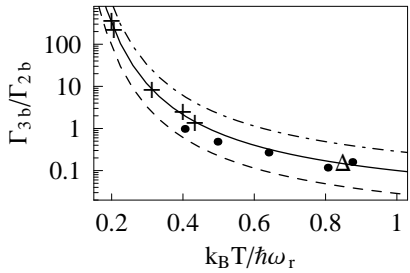,width=0.7 \columnwidth}
\caption{Ratio between the scattering rates for the two routes to 
thermalization and breakdown of integrability in 1D systems: 
$\Gamma _{2b}$ for two-body collisions leading to excited 
transverse states, and $\Gamma _{3b}$ for the effctive three-body 
collisions. Units on the axes are dimensionless.
$\zeta =$ 0.002 (dashed curve), 0.007 (solid curve), and 0.02
(dot-dashed curve). The points represent the predicted ratios
for various sets of experimental parameters from \cite{js1}
(points), \cite{js2} (crosses), and \cite{va} (triangle). }
\label{fig:ScatRat}
\end{figure}

Using Eq. (\ref{eq:11}) we then obtain the collision rate for the
process shown in Fig.~\ref{fig:1}(a):
\begin{eqnarray}
    \Gamma_{3b}&=&C_{3b}\frac {\hbar n_{1D}^2}m
    \left( \frac {\alpha _s}{l_r} \right)^4
    = C_{3b} \, \omega_r \, \zeta^2,
    \label{eq:17b}
\end{eqnarray}
with $C_{3b}=\frac {72\xi ^2}{\sqrt{3} \pi^2} \approx 5.57$.
Comparison to the two-body rates for typical experiments are given
in Fig.~\ref{fig:ScatRat}.

The result of Eq. (\ref{eq:17b}) may seem counterintuitive at
first: the collision rate is independent of temperature, and is
proportional to $\zeta^2$ and the radial confinement $\omega_r$.

The physics behind the first observation is related to the fact
that the collision kinetic energy is small compared to the virtual
excitation energy (according to assumption Eq.~(\ref{eq:1})).
Consequently the composite matrix element of the second-order
process should not depend (in leading order) on the velocities of
colliding particles and hence on temperature (see
Eq.~(\ref{eq:5})). In addition the phase space volume for the
scattered particles is independent on the incoming momenta $k_1$,
$k_2$, and $k_3$.

Since effective three-body elastic scattering is the dominant
process the scattering rate must be proportional to the 3D density squared,
$(n_{1D}/w^2_r)^2$.  On the other hand, the scattering rate
contains the square of the matrix element corresponding to the
diagram in Fig.~1(a), where each vertex is proportional to $\alpha
_s$, therefore this rate is proportional to $\alpha _s^4$. The
factor $\hbar/m$ provides the correct dimensionality (s$^{-1}$).

We can now compare the scattering rates for the two routes for
thermalization and breakdown of integrability in 1D systems:
thermally excited two-body collisions $\Gamma_{2b}$
(Eq.~\ref{eq:20}) or effective three-body collisions $\Gamma_{3b}$
(Eq.~\ref{eq:17b}). For $k_BT<\hbar \omega _r$ we find a simple scaling:
\begin{equation}\label{eq:ScatRat}
    \frac{\Gamma_{3b}}{\Gamma_{2b}} \approx \frac{36}{\sqrt{6} \pi^2} \;
        \zeta \, e^{\frac {2\hbar \omega _r}{k_BT}} \approx
        1.97 \; {\zeta}{e^{ \frac {2\hbar \omega _r}{k_BT}}}  . 
\end{equation}
For large $\zeta $ and small temperatures the scattering rate due
to {\em virtual} excitations dominates, and can violate
integrability even when thermalization processes due to simple
two-body collisions are frozen out. A  detailed comparison of the 
two rates and their relation to
experimental parameters is given in Fig. ~\ref{fig:ScatRat}.
The scattering rate due to {\em virtual}
excitations of the radial modes can dominate over real
excitations for typical parameters of the recent experiment \cite{js2}.

The damping rate of an elementary excitation that moves in a
{\em degenerate} Bose gas much faster than the speed of sound also has
the functional form of Eq. (\ref{eq:17b}), with $C_{3b}$ being
replaced now by $C_{fast}\approx 13.76$. This result 
suggests that thermalization is very similar also in the largely
unexplored quantum decoherent regime \cite{IBc} of quasi-1D
bosonic dynamics.

To quantify the thermalization, and thereby the violation of
integrability in a quasi-1D system by the effective interaction
(\ref{eq:11}), we consider a non-degenerate, weakly-interacting
(the Lieb-Liniger parameter \cite{LL} $\gamma =2\alpha
_s/(n_{1D}l_r^2)$ being much less than 1) gas of bosonic atoms
\cite{foot1} and write out the Boltzmann equation with a
three-body collision integral \cite{beq3b}, taking into account
the indistinguishability of the particles:
\begin{eqnarray}
    \frac d{d t} f_k&=&\frac {72\xi ^2\omega _r^2
    \alpha _s^4m}{\sqrt{3}\pi ^3\hbar }\int _{-K}^K
\frac {dq}{\sqrt{K^2-q^2}} \int _{-\infty }^\infty dk^\prime
    \int _{-\infty }^\infty dk^{\prime \prime } \times \nonumber \\
    &&\left( f_{k_0-q}f_{k_+}f_{k_-}-f_k f_{k^\prime }f_{k^{\prime \prime }}
\right) , \label{eq:12} \\
    k_0&=&\frac {k+k^\prime +k^{\prime \prime }}3,  \quad 
    k_\pm ={k}_0 +\frac q2\pm \frac {\sqrt{3}}2\sqrt{K^2-q^2} ,
\nonumber \\
K&=&\frac 23 \sqrt{k^2+k^{\prime \, 2}+k^{\prime \prime \, 2}- kk^\prime
-kk^{\prime \prime }-k^\prime k^{\prime \prime }}.
\end{eqnarray}

\begin{figure}
 \epsfig{file=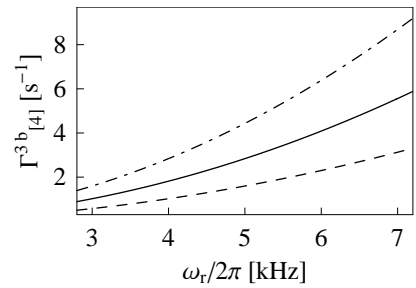,width=0.7\columnwidth}
\caption{Dependence of the rate $\Gamma _{[4]}^{3b}$ of
thermalization induced by effective three-body collisions on the
radial trapping frequency. $n_{1D} = 30~\mu $m$^{-1}$ (dashed
curve), $40~\mu $m$^{-1}$ (solid curve), and $50~\mu $m$^{-1}$
(dot-dashed curve). } \label{fig:Thermalization}
\end{figure}

To solve Eq.~(\ref{eq:12}) we use the following {\em ansatz} for
the perturbed momentum distribution
\begin{eqnarray}
f_k(t)= \frac {n_{1D}}{\sqrt{\pi }k_{th}} \exp (-k^2/k_{th}^2)
[ 1+\varepsilon _4(t)H_4 (k/k_{th})] , \label{eq:16}
\end{eqnarray}
where $k_{th}={\sqrt{2mk_BT}/\hbar }$ and 
$H_4$ is the Hermite polynomial of the 4th order. We choose
this form, since it is the simplest nontrivial perturbation that
retains $\int dk\, kf_k=0$. Linearizing Eq.~(\ref{eq:12}) with
respect to the perturbation amplitude $\varepsilon _4(t)$ we
obtain the exponential solution $\varepsilon _4(t)=\varepsilon
_4(0) \exp (-\Gamma _{[4]}^{3b}t)$ with the decrement 
\begin{eqnarray}
    \Gamma _{[4]}^{3b}&=&C_{[4]}\frac {\hbar n_{1D}^2}m\left(
    \frac {\alpha _s}{l_r} \right) ^4= C_{[4]} \, \omega_r \, \zeta^2
    \label{eq:17}
\end{eqnarray}
where the numerical constant $C_{[4]}=\frac {64\xi ^2}{3\sqrt{3}\pi
^2} \approx 1.65$. Taking the perturbation proportional to a
higher-order Hermite polynomial $H_n$ leads only to a minor the
modification of the numerical prefactor, leaving the functional
dependence on the parameters of the system unchanged (for example,
$n=5$ or 6 increases the thermalization rate by the factor $5/4$
or $13/9$, respectively). Fig.~\ref{fig:Thermalization} shows
numerical values of $\Gamma _{[4]}^{3b}$ as a function of the 1D
density of $^{87}$Rb atoms and the radial trapping frequency.

Similarly, we calculate numerically the thermalization rate
$\Gamma _{[4]}^{2b} $ for two-body collisions involving the {\em
real } transitions between the ground and excited radial states.
The velocity distribution of atoms in the ground and excited state
was perturbed in the same way, as given by Eq. (\ref{eq:16}), the
Boltzmannian distribution of overall populations between the
levels being kept intact. In the parameter range of interest we find 
numerically $\Gamma _{[4]}^{2b}\approx (0.33\pm 0.03)\Gamma _{2b}$, i.e.
\begin{equation}
 \Gamma _{[4]}^{2b} \approx 0.93\,
 \omega _r \zeta e^{-\frac{2\hbar \omega _r}{k_BT} }.
 \label{eq:30}
\end{equation}
The ratio of the thermalization rates for the two-body and
three-body processes is very close to the respective ratio of the
collision rates, shown in Fig.~\ref{fig:ScatRat}.

It is interesting to note that we find for both processes, 
the two-body collisions to real transverse states and the effective 
three-body processes via virtual excited states, that thermalization in
1D needs about 3 collisions which are able to distribute energy.
This is very close to the 2.7 collisions required for
thermalization in 3D \cite{WuFoot}.

For the typical parameters of an ultracold $^{87}$Rb gas on an atom chip
\cite{js2} ($\omega _r\approx 2\pi \times 3$~kHz, $n_{1D}\approx
50~\mu $m$^{-1}$) we obtain $\Gamma _{[4]}^{3b}\approx 2$~s$^{-1}$.
This thermalization rate is temperature-independent and much
larger than the one calculated from the simple two-body collisions
with the energy sufficient to excite radial modes
$\Gamma _{[4]}^{2b}\approx 3 \times 10^{-3}$~s$^{-1}$ at the lowest
temperatures measured (30 nK).  The estimated $\Gamma _{[4]}^{3b}$
is consistant with the time needed for evaporative cooling of a
$^{87}$Rb gas on an atom chip well below $\hbar \omega _r$
\cite{js1,js2}.

Comparison of our thermalization rates to the experimental results
of Kinoshita, Wenger and Weiss \cite{dw1} is more qualitative. In
this experiment, a degenerate $^{87}$Rb gas in a two-dimensional
optical lattice ($\omega _r\approx 2\pi \times 67$~kHz,
$n_{1D}\approx 10~\mu $m$^{-1}$) is split by a laser Bragg pulse
into two groups with opposite kinetic momenta, which begin to
oscillate in a weak trapping potential in $z$-direction, colliding
each half-period of the oscillation. The experiment is close to
the strongly interacting regime ($\gamma \sim 1$), and one has to
take the strong suppression of three-body collisions into account.
The rate of relaxation of the relative motion of the two groups of
atoms can be written as
\begin{equation}
    \Gamma _{rm}=C_{rm}\eta _\tau  \, \omega_r \, \zeta^2 {\cal F}(\gamma) ,
    \label{eq:18}
\end{equation}
where $C_{rm}\sim C_{3b}$ and $\eta _\tau <1$ is the fraction of
the period, when the two groups of atoms with opposite velocities
overlap in space. The dominant factor in
Eq.~(\ref{eq:18}) is the suppression of three-body scattering  ${\cal
F}(\gamma)$ via atom-atom correlations in both initial and final
states. The experiment \cite{fa1} showed 7-fold suppression of
three-body {\em inelastic} losses in a quasi-1D bosonic gas with $\gamma
\approx 0.5$ compared to a regular BEC. However, there is a
difference between elastic scattering discussed here and that of
Ref. \cite{fa1} where the final state corresponds to a molecule and
the fast atom carrying away the energy and momentum released in
the three-body recombination process. In the {\em elastic} scattering
related to dissipation the suppression enters in both the incoming
and the outgoing channel and should be the square of the measured
value. Reading the predictions for the parameters in \cite{dw1}
($\gamma \approx 1.4$) from the estimates in \cite{fa1} we find a
suppression factor ${\cal F}(1.4) \sim 4 \cdot 10^{-4}$ and
$\Gamma _{rm}\lesssim 0.03$~s$^{-1}$, which is consistent with the
experimentally obtained lower bound to the relative motion damping
rate $\Gamma _{rm}^{-1}> 20$~s.

The mechanism discussed here which leads to the breakdown of
integrability in 1D is to a certain extent similar to the virtual
association of atoms to a molecular dimer \cite{yurbro}. In the
present discussion, virtual excitation of radial modes during a
two-atom collision temporarily localize the interatomic distance
on the length scale $\sim l_r$.  Scattering a third atom on such a
transient structure of finite size and  mass $2m$ leads to
thermalization and violates integrability. In Ref. \cite{yurbro}, 
collisions of a third atom bring ``virtual'' dimers, enhanced in
size by a Feshbach resonance down to the energy shell, thus
bringing about ``quantum chemistry'' in 1D.

To summarize, we identify effective three-body collisions which
arise in the second order of perturbation theory as a 
mechanism of 
breakdown of integrability and, hence, thermalization in 1D atomic
gases.  These processes are associated with virtual excitation of
radial modes and for weakly interacting quasi-1D Bose gases can be
dominant at $k_BT<\hbar \omega_r$. Our estimations of the relaxation
rates in are consistent with recent experimental observations
\cite{js1,js2,dw1}.

This work is supported by the EC (STREP MIDAS) the INTAS
and the FWF. I.E.M. acknowledges the Lise Meitner Fellowship (FWF).


\begin{thebibliography}{99}
\bibitem{foot2}
The effective 1D coupling constant of pairwise interacting bosonic
atoms in a waveguide changes with confinement as $g_{1D}=2\hbar
\omega _r \alpha _s /[1-\frac{C^\prime \alpha _s}{\sqrt{2}l_r}]$,
$C^\prime \approx 1.46$ \cite{ol1} due to virtual excitation of
the radial modes and increases as $\alpha _s/l_r$ grows. This
points to the ratio $\alpha _s/l_r<1$ as the measure how much the
1D approximation is violated.
\end{thebibliography}
\end{document}